Social Network Analysis for Social Neuroscientists

Elisa C. Baek[1,2], Mason A. Porter[2] & Carolyn Parkinson[1,3]

[1]Department of Psychology, University of California, Los Angeles

[2]Department of Mathematics, University of California, Los Angeles

[1,3]Brain Research Institute, University of California, Los Angeles

Author Note

This work was supported by NSF grant SBE-1835239 and by NSF SBE Postdoctoral Research Fellowship grant 1911783.

Correspondence concerning this article should be addressed to Elisa C. Baek, Department of Psychology, University of California, Los Angeles, Los Angeles, CA 90095. Contact: elisabaek@ucla.edu

Word count (main text excluding footnotes, tables, boxes, figures, figure legends, and references): 9,901

Abstract

Although social neuroscience is concerned with understanding how the brain interacts with its social environment, prevailing research in the field has primarily considered the human brain in isolation, deprived of its rich social context. Emerging work in social neuroscience that leverages tools from network analysis has begun to pursue this issue, advancing knowledge of how the human brain influences and is influenced by the structures of its social environment. In this paper, we provide an overview of key theory and methods in network analysis (especially for social systems) as an introduction for social neuroscientists who are interested in relating individual cognition to the structures of an individual's social environments. We also highlight some exciting new work as examples of how to productively use these tools to investigate questions of relevance to social neuroscientists. We include tutorials to help with practical implementation of the concepts that we discuss. We conclude by highlighting a broad range of exciting research opportunities for social neuroscientists who are interested in using network analysis to study social systems.

Social Network Analysis for Social Neuroscientists

Humans are social beings and are immersed in intricate social structures. Social interactions and relationships play important roles in healthy human development and functioning (House et al., 1988; Seeman, 1996; Uchino, 2006), and the need to navigate complicated social interactions for survival advantage may have contributed to human brain evolution (Dunbar, 2008). Nevertheless, most work in social neuroscience has studied individual cognition in isolation, deprived of its rich social context. As demonstrated recently (Morelli et al., 2018; O'Donnell et al., 2017; Parkinson et al., 2017, 2018; Zerubavel et al., 2015), social neuroscientists can leverage tools from network analysis to characterize the structure of individuals' social worlds to improve understanding of how individual brains shape and are shaped by their social networks (Weaverdyck & Parkinson, 2018).

Recent work that relates characteristics of individuals' social networks to their behaviors and attitudes has uncovered important insights into how people are impacted by the structures of their social world. For instance, one study that used network tools to characterize the patterning of relationships in an organization showed that individuals who are not well-connected to well-connected others are especially likely to be the object of negative gossip and scapegoating (Ellwardt et al., 2012). As this example and other recent research demonstrate, the features of an individual's social network can profoundly impact how they feel (Coviello et al., 2014; Fowler & Christakis, 2008); how they behave toward others (Ellwardt et al., 2012; Paluck & Shepherd, 2012; Shepherd & Paluck, 2015); and their general behaviors, attitudes, and ways of seeing the world (Aral & Walker, 2012; Centola, 2011; Christakis & Fowler, 2007; Oh & Kilduff, 2008). Clearly, social network attributes significantly influence individuals' cognition, behavior, and affect. However, the mechanisms that underlie these effects remain poorly understood. In this paper, we provide an overview of key theory and methods in network analysis (especially for social systems) and discuss practical examples to highlight how network analysis can be useful for social neuroscientists who are interested in relating individual cognition to the structure of social environments. We also include two tutorials to help with practical implementations of the concepts in this paper.

## Key Concepts of Network Analysis for Social Systems

We now introduce some key concepts of network analysis that are particularly relevant for understanding social systems (see also Table 1).

### Nodes and Edges

Suppose that we want to characterize how people are connected to one another in a small town. What do we want to know? We may first wish to identify the individuals in the network. We represent individuals in a network (i.e., "graph") by *nodes*, which are often called "vertices" in mathematics and "actors" in the context of social systems (see Figure 1). For introductions to networks, see Wasserman & Faust (1994) for a sociological perspective, Kolaczyk (2009) for a statistical perspective, and Newman (2018) for a physical-science perspective). In our hypothetical example, a node may represent an inhabitant of a town. We may next wish to understand who is connected to whom in a network. Considering such connections is what differentiates studying a group (a collection of nodes) from a network (which also encodes the connections between nodes). We represent these connections by edges (which are often called "ties" or "links"). Depending on the questions of interest, edges can encode different relationships. For instance, edges can represent friendship (e.g., in academic cohorts; Parkinson et al., 2017, or in student organizations; Zerubavel et al., 2015) or professional relationships (e.g., in sports teams; Grund, 2012, or in private firms; Zaheer & Bell, 2005). One can define such relationships in terms of subjective reports (e.g., of who likes whom; Zerubavel et al., 2015, or who trusts whom; Morelli et al., 2018) or the frequency of particular types of interactions or communications (e.g., physical encounters; Read, Eames, & Edmunds, 2008, or exchange of e-mails; Wuchty & Uzzi, 2011). Edges can also represent other phenomena, such as shared attributes (e.g., attendance at the same social events; Davis,

Gardner, & Gardner, 1941) or common behavioral patterns (e.g., voting similarities; Waugh, Pei, Fowler, Mucha, & Porter, 2009).

It is sometimes important to consider the directions of edges. For example, in a friendship network, we may place an edge from node A to node B if A reports "liking" B; however, although it may be awkward, it is possible that B may not "like" A. One can represent such relationships with directed edges, with an arrow pointing from one node to another (for example, from A to B). In other cases, edges are undirected, either because the criterion that is used to define them is inherently undirected (e.g., shared attributes) or because it can sometimes be pragmatic to consider edges as undirected. For example, a researcher may choose to consider an undirected "friendship" tie between A and B if and only if they both reported liking one another to impose a stringent definition of friendship and/or if the researcher wishes to relate these data to other undirected data, such as interpersonal similarities. It is also sometimes desirable to consider edge weights to represent relationship strengths. For example, one can encode interaction frequency with edges that are weighted by the number of interactions (during some time period) between two actors. In other cases, edges are unweighted, either because one obtains them in a way that is unweighted by nature (e.g., edges that encode the existence of a relationship), or because there is a compelling reason to consider edges as unweighted. For example, to characterize only meaningful relationships, one may choose to use an edge that represents a relationship between two people if and only if it meets or exceeds a minimum threshold on the number of interactions.

In summary, edges in a network can be directed or undirected, and they can be either weighted or unweighted. Choosing how much information to include in edges depends both on how data are acquired (e.g., by asking questions that produce binary or continuous responses) and on how they are encoded in a network (e.g., decisions to threshold and binarize continuous responses). There are advantages and disadvantages to using directed and weighted edges, rather than using edges that are undirected and unweighted. Although directed and weighted edges can provide additional information about the nature of a relationship between two nodes, they can also complicate analysis. As we discuss in the following sections, they can complicate the characterization of various network measures and affect associated inferences. (Some methods also do not work in such more complicated cases; Newman, 2018.) Consequently, researchers should carefully consider these factors when deciding how to represent a social network. Other complications are that a network can include multiple types of edges ("multiplex networks") and the nodes and edges in a network can change over time ("temporal networks"). We discuss these issues later (see the section on "Multilayer Networks"), and they are reviewed in detail elsewhere (Aleta & Moreno, 2019; Kivelä et al., 2014).

**Sociocentric Networks versus Egocentric Networks**

One can study networks either by considering a sociocentric network (which is also called a "complete network"; Marsden, 2002; Newman, 2018) or by taking an egocentric (i.e., "ego-network") approach (Crossley et al., 2015). A sociocentric-network approach encapsulates a complete picture of who is connected[1] with whom in a network. One can construct a sociocentric social network by asking each person in a network about those with whom they are connected directly using a desired type of connection (depending on the question of interest). For instance, one might survey all members of a sports team to characterize a friendship network by asking who their friends are or who they turn to for emotional support. Recent work in social neuroscience that leverages tools from network science has often used a sociocentric-network approach to characterize relatively small, bounded networks. Bounded networks (which are also called "closed networks") have clearly defined boundaries. In the strictest adherence to the definition of "bounded," the boundaries of a social network are known perfectly, because individuals reside in a restrictive physical environment, such as a remote island (Brent et al., 2017), or are assigned to isolated social groups (Sallet et al., 2011). It is difficult to obtain perfectly bounded networks in humans, but recent work in social neuroscience has characterized relatively bounded networks, such as academic programs, dorms, and clubs (Morelli et al., 2018; Parkinson et al., 2017, 2018; Zerubavel et al., 2015). It has then collected neuroimaging data from some of the members of these relatively bounded networks to relate neural processing to social network measures. Such an approach demonstrates one useful way to study individual cognition in the context of a broader social environment.

---

[1] We use the term "connected" to indicate that two individuals have a relationship with one another. We use the term "connected directly" to indicate that two individuals are connected with a distance of 1 (i.e., they are "adjacent" to each other in a network). Our use of the term "connected directly" is synonymous with "direct ties" and the mathematical definition of "adjacent." We also use the term "connected indirectly" to indicate that two individuals do not have a direct relationship with one another, but they each have relationships through one or more third parties (e.g., through mutual friends). We use the term "connected" throughout the paper, because we expect this terminology to be intuitive to our target audience for conveying our intended meaning. It is important not to confuse our usage of "connected" with the use of it to describe graphs or components of graphs (rather than individual nodes) in graph theory. The latter usage of "connected" refers to the idea that a path exists between every pair of nodes in a graph or in a component of a graph (Newman, 2018).

It is often insightful to study social networks using an ego-network[2] (i.e., egocentric-network) approach. An ego network is a network based on an individual (the "ego") and their friends (the "alters"). One can construct ego networks in a few different ways. If one possesses data on an entire bounded network, one can use it to extract "objective" ego networks that consist of one individual and their friends. In such cases, where one obtains ego networks as part of a study that also characterizes sociocentric networks, researchers may also be interested in comparing an individual's perceptions of a network to actual characteristics of the network. Such a comparison can lead to interesting questions about how people think about their relationships and relate to the social world around them through "cognitive social structures" (Krackhardt, 1987). In this case, one can construct "subjective" ego networks by asking individuals ("egos") to complete a questionnaire about the people ("alters") to whom they are connected[1] directly and whether these people are also connected directly to one another. For instance, one can survey a single member of a sports team to ask who their friends are and which of their friends are also friends with one another. Although it is relatively uncommon to obtain data on individuals' perceptions of relationships between third parties in situations when one already has characterized a sociocentric network with those individuals (and their alters), such an approach provides a useful way to explore questions about individuals' perceptions of their networks and characteristics of a sociocentric network.

---

[2] By default, an ego network is a 1-ego network, which consists of an ego's alters and the edges between those alters. A 1-ego network thereby consists of the nodes and edges that are in an ego's personal social network (Crossley et al., 2015; Jeub, Balachandran, Porter, Mucha, & Mahoney, 2015). When mathematically analyzing 1-ego networks, one often does not include the direct connections between the ego and the alter, as one instead concentrates on the direct connections that exist between the alters. When we write "ego networks", we refer specifically to 1-ego networks. One can go further than an ego's 1-neighborhood by obtaining information about the alters' additional connections, beyond just those who have direct ties with an ego (e.g., by also obtaining the ego networks of each of the ego's alters). This yields a 2-ego network, which gives information about the nodes of distance 2 or less from an ego (e.g., "friends of friends" of the ego). One can iterate this process further to obtain *k*-ego networks (i.e., about all nodes within distance *k* of an ego) and thereby encode information about larger social structures in which an ego is immersed. A benefit of *k*-ego networks is that they provide more information about the broader social contexts of an individual than 1-ego networks, although it is often more cumbersome to obtain them in practice.

It is most common to obtain and characterize ego networks independently, without possessing information about an associated sociocentric network. In this situation, one typically characterizes ego networks through questionnaires that ask one individual (the "ego") about the people (the "alters") to whom they are connected directly and, in some cases, whether those people are connected directly to one another. When obtaining a sociocentric network is infeasible or inconvenient, employing an ego-network approach alone can be useful. However, ego networks do not provide a complete picture of an entire sociocentric network, which limits the type of inferences that one can draw from such data. For instance, when using an ego-network approach, if one finds that individual differences in network position[3] are associated with a behavioral or neural outcome, it is unclear whether this relationship is due to actual differences in network position or differences in individuals' perceptions of their network position (e.g., how many friends people think that they have versus how many friends they actually have). Despite their limitations, a key advantage of ego networks over sociocentric networks is that it is much easier to collect the former, and one can conveniently add them to a study by administering questionnaires to individuals in isolation. Several new insights in social neuroscience have resulted from the use of ego-network approaches. For example, estimates of the number of connections between the ego and other people from self-reporting and Facebook ego networks are associated with structural and functional differences in brain regions (Hampton et al., 2016; Von Der Heide et al., 2014), and individual differences in network position that were identified from Facebook ego networks were associated with brain activity during a social-influence task (O'Donnell et al., 2017).

---

[3] We use "network position" as a general term to refer to features that are related to an individual's location in a social network (e.g., with whom they are close in social ties) and their node-level characteristics (e.g., centrality measures that quantify the influence of an individual in relation to other individuals). It is important not to confuse our use of this terminology with the more specific use of "network position" in relation to positional analysis (Wasserman & Faust, 1994).

**Mathematical Representation of Networks**

One can represent a network mathematically using an adjacency matrix[4]. An adjacency matrix **A** of a network is an $n \times n$ matrix (where $n$ is the number of nodes) with elements $A_{ij}$. In an undirected and unweighted network, $A_{ij}$ is 1 if there is an edge between nodes $i$ and $j$, and $A_{ij}$ is 0 if there is no edge between nodes $i$ and $j$. Because $A_{ij} = A_{ji}$ in an undirected network, an adjacency matrix of such a network is symmetric (see Figure 1). One can also represent a network using an edge list, which enumerates node pairs that are connected directly by edges (see Figure 1).

**Social Distance**

Consider two strangers who are meeting for the first time. After speaking with one another for a while, they may be surprised to find that they have an acquaintance in common and then marvel at how small the world seems to be. Anecdotal evidence suggests that many people have had this sort of experience, reflecting the "small-world phenomenon" (i.e., the idea that people in general are connected to each other by relatively short chains of relationships; Newman, 2018). Many people have an intuitive sense of the small-world phenomenon, but one may wonder how "small" the world really is (i.e., how close together, in terms of social ties, people actually are). In his pioneering studies of social distance, social psychologist Stanley Milgram and colleagues sought to test this question (Milgram, 1967, 1969). In these experiments, they recruited participants in the midwestern part of the United States and instructed them that their goal was to send a package (which included an official-looking letter and a stack of cards that was meant to track each person in the chain) to reach a target individual in Massachusetts. If they did not personally know the person on a first-name basis, they were instructed to forward the package to one of their direct connections who they thought was likely to be closer to the target. Milgram and his colleagues found that, on average, it took six steps for the packages (among those that completed their journey) to reach the target individual (see Figure 2). This finding has been popularized in popular culture as "six degrees of separation," expressing the idea that any two people in the world are separated by six or fewer social connections. More recently, scholars have examined the small-world phenomenon through algorithmic frameworks (Kleinberg, 2000, 2011) and experiments like those of Milgram and colleagues have been conducted using communication channels such as e-mail (Dodds et al., 2003) and online social networks (Ugander et al., 2011).

---

[4] More complicated network structures, such as multilayer networks, have more complicated adjacency structures (Kivelä et al., 2014).

In this section, we overview concepts and methods for calculating social distance and discuss their utility for examining questions of interest to social neuroscientists. Given a network of nodes and edges, one can calculate a distance between two nodes (e.g., how far A is from B in a network). There are several ways of calculating distances in a network. The simplest is geodesic distance, which is the smallest number of edges that one needs to traverse to connect two nodes in a network (i.e., a shortest path). Two nodes can be connected either by direct ties (e.g., "friends" in a friendship network, with a distance of 1, because they are "adjacent" in the network) or by indirect ties (e.g., "friends of friends", which yields a distance of 2, "friends of friends of friends", which yields a distance of 3, and so on). The numerical values of social distance lead to different sociological inferences, which depend on context. For instance, consider the friendship network of a first-year cohort at a university. Suppose that A and B in this network are separated by a social distance of 4 (e.g., "friends of friends of friends of friends"). We may be interested in interpreting the absence of friendship between these two actors based on the social distance of 4. Perhaps they are distant from one another because they do not have much in common with each other. However, we would make different inferences from this social distance of 4 depending on whether the two individuals live in dorms on opposite sides of the university campus or on the same floor of the same dorm. In the first scenario, the two individuals may be distant from one another in friendship ties due to a lack of opportunity to interact (and not necessarily a lack of common interests). By contrast, the two individuals in the second scenario likely have had opportunities to interact but are not friends, so a lack of common interests may be a more plausible explanation for the large social distance between them. Missing data can also complicate the interpretation of social distance, as missing ties can lead to an overestimation of distance between two individuals. For example, in this scenario, if we are missing data from an individual in the network who is friends with both individuals (but we know that these two individuals are definitely not friends with each other), the actual distance between the two individuals is 2, rather than 4. Therefore, when drawing inferences from social distance, it is advantageous to choose networks that are bounded (so we do not miss indirect connections between individuals, as this may lead to overestimation of some social distances) and where we can be confident that opportunities to interact are relatively equally distributed across the network (to constrain interpretations of the potential causes of the relative distances between people). That said, the reason that actors are distant from each other may not matter as much in other situations, such as when characterizing the spread of information or behavior. When considering which network measures to use, researchers should ensure that they use methods and tools that are appropriate for their questions of interest.

Recent neuroimaging work suggests both that the human brain tracks the social distance between oneself and familiar others and that people spontaneously retrieve information about others' social-network positions when viewing their faces (Morelli et al., 2018; Parkinson et al., 2017; Zerubavel et al., 2015). This spontaneous retrieval of social-network knowledge when encountering familiar others may help people respond appropriately when interacting with different people. There is also evidence that the brain not only tracks information about social-network position, but also that it influences and is influenced by a person's social networks. For example, friendship is associated with similarity of neural responses to naturalistic stimuli. Recent work found that participants tend to have more similar time series of neural responses to audiovisual movies to people with whom they are connected directly (e.g., friends) than to people with whom they are only connected indirectly (e.g., friends of friends), with neural similarity decreasing with increasing social distance (Hyon et al., 2020; Parkinson et al., 2018). This suggests that (1) people process information about the world in similar ways to those who are socially close to them and that (2) individual brains may shape, and be shaped by, other brains that surround them. Such results demonstrate that one can leverage tools from network analysis to advance understanding of how individual brains represent and process the world around them.

**Distance in Weighted Networks.** Thus far, we have focused our discussion on geodesic distance, which is the simplest way of computing distance and is used often in unweighted networks. Computing distance in weighted networks is more complicated, and there are many ways to do it. A comprehensive discussion is beyond the scope of this paper, but see Cherkassky, Goldberg, and Radzik (1996) for a detailed consideration of shortest paths in weighted networks. A common way to calculate distance in weighted networks is to convert pairwise weights to costs and then use Dijkstra's shortest-path-first algorithm (Dijkstra, 1959; Newman, 2001). See Box 1 for an overview of Dijkstra's algorithm and important considerations for interpreting distances in weighted social networks.

**Centrality**

It is often of interest to characterize the importance of actors (or of edges between them) in a social network. For instance, we may wish to know who is well-connected or popular in a school. The concept of "centrality" in network analysis is helpful for examining such questions (Newman, 2018). There are myriad variants of centrality; we discuss some of the most common types in network analysis of social structures, with a focus on methods for calculating these centrality measures in unweighted, undirected networks. We also point to some resources for discussions of variations of these measures in weighted and directed networks. See Bringmann et al. (2019) for important caveats about studying and interpreting centralities in networks.

**Degree Centrality.** Degree centrality (i.e., "degree") is the number of edges that are attached to a node, so it is the number of direct connections of a person in a social network (see Figure 3). Another way to think about degree is in terms of "walks" across edges in a network. Consider a robot that is walking around a social network. Given an undirected and unweighted network, we calculate the degree of a node by taking the number of different ways that the robot can reach that node via a walk length of 1 (i.e., from a directly connected neighbor). Although degree is a simple concept to grasp intuitively without illustrating it with a walking robot, we include this description because it is helpful for comparing degree to other centrality measures. There are various generalizations of degree that incorporate edge directions and/or weights, and we discuss some of them in the "Consideration of Direction and Weights in Centrality Measures" section.

**Eigenvector Centrality.** Although degree is a useful measure of centrality, it counts the number of connections of a node without considering the quality of those connections. Consider a townsperson who does not have many friends but is friends with the mayor, who has a large degree (and hence is well-connected in that respect). Although that townsperson has few friends, they may have more influence in the town than an individual with many friends with small degrees. Eigenvector centrality takes this type of connectivity into account, providing one way (see Figure 3) to capture how well-connected a person is to other well-connected people (Bonacich, 1972). One calculates the eigenvector centralities of the nodes in a "connected" (in the graph-theoretic sense) network as the components of the leading eigenvector of the network's adjacency matrix[5] **A**. Another way to visualize the idea behind eigenvector centrality is through a random walk. Suppose that a robot goes on an infinitely long random walk through a network. The eigenvector centrality of a node is proportional to the frequency of visits by the robot during its walk in the network. The robot visits a node with a large eigenvector centrality more often than a node with a small eigenvector centrality, because the former node's direct neighbors are well-connected to other nodes in the network. Using this idea, one can derive the formula for eigenvector centrality using a random walk, and different variants of random walks lead to different types of eigenvector-based centralities (Masuda, Porter, & Lambiotte, 2017).

Eigenvector centrality has been associated with various social and health-relevant phenomena in humans—including happiness (Fowler & Christakis, 2008), body weight (Christakis & Fowler, 2007), and job retention (Ballinger et al., 2016)—and with reproductive success in animals (Brent, 2015), suggesting that indirect ties (e.g., friends of friends, friends of friends of friends, and so on) may influence an individual's well-being and behavior (and vice versa). Additionally, people may be more likely to know who is well-connected to well-connected others than who has a lot of friends. For instance, in a large school, people may be keenly aware of which individuals are popular in a popular group, but they may be less aware of which individuals in a less-popular group have many friends. This knowledge of who is well-connected to well-connected others has important implications. Mistreating an individual who is well-connected with well-connected ties may be risky, as the individual may be defended by their friends and their friends of friends, whereas mistreating a poorly connected individual may have minimal consequences, given their limited influence (Ellwardt et al., 2012; Salmivalli et al., 1996). In light of these scenarios, eigenvector centrality may be particularly useful when studying how people perceive social status in a network and how these perceptions shape behavior. For a brief discussion of PageRank, a variation of eigenvector centrality, see our Supplementary Material.

---

[5] As we described earlier[1], a network is "connected" in this sense if, for all pairs of actors, there is a walk between these actors. A directed network where one can reach any node by a path that starts from any other node is called "strongly connected".

**Diffusion centrality**. Diffusion centrality, which generalizes both eigenvector centrality and Katz centrality (another notion of importance that is based on a walk on a network; Newman 2018), captures an individual's centrality with respect to a simple spreading process on a network (Banerjee et al., 2013). Calculating diffusion centrality may be useful for social neuroscientists who are interested in characterizing how central individuals are in their ability to spread items (such as information) in a dissemination process. Prior work has suggested that people are accurate at identifying others who are good at spreading information in a social network and that these estimates are correlated with diffusion centrality (Banerjee et al., 2014).

**Betweenness Centrality**. Another type of centrality is geodesic betweenness centrality, which measures the extent to which shortest paths (or, in generalizations of betweenness, other types of short paths) between pairs of nodes pass through a node. Suppose that a robot is traversing a network and takes a shortest path between each pair of nodes. One can calculate betweenness centrality of a node by tracking the number of times that the robot passes through the node to connect each pair of nodes (see Figure 3). It is common to interpret betweenness centrality as a measure of brokerage, because it captures some information about the extent to which a node connects distally connected nodes (Wasserman & Faust, 1994). For instance, an individual with a large betweenness centrality may have a high capacity for brokerage, because more of their friends have to go through them to communicate with one another. However, one should be cautious when interpreting betweenness as a measure of brokerage, as many different factors in network structure (including ones that are unrelated to a given individual) can strongly influence betweenness (Everett & Valente, 2016). In large networks, for instance, an individual may not be well-connected (as quantified, e.g., by a small degree) and not well-connected to well-connected others (as quantified, e.g., by a small eigenvector centrality), but they may still have a large betweenness. This individual may be in the periphery of multiple groups of friends; although they may broker information between groups of otherwise unconnected nodes (e.g., two friendship groups), they may not be very influential in either of the individual groups. Another possibility is that individuals may have a large betweenness if they are connected directly to nodes that are brokers, even if they are not much of a broker themselves. If a researcher is interested in characterizing individual differences in socio-behavioral tendencies that are related to brokerage (e.g., how often people introduce their friends to one another), it may be useful to calculate local network measures (such as local clustering coefficient; Watts & Strogatz, 1998, and constraint; Burt, 2004). Similar to many centrality measures, betweenness is not robust to noise in data (e.g., missing edges), so it is necessary to pay careful attention to such issues (Bringmann et al., 2019; Everett & Valente, 2016).

**Considering Edge Directions and Weights in Centrality Measures**. In directed networks, each node has both an in-degree centrality (the number of edges that point to it) and an out-degree centrality (the number of edges that emanate from it). Depending on the question of interest, it may be appropriate to calculate versions of centrality measures for networks with directions and/or weights. In some cases, generalizations are straightforward; for example, generalizing betweenness centrality to directed networks only requires restricting the node pairs (i.e., origin–destination pairs) that one considers, and one can directly generalize eigenvector centrality to weighted and directed networks by defining it based on random walks or as the leading eigenvector of an adjacency matrix. PageRank (see Supplementary Material) is formulated specifically for directed networks and generalizes to weighted networks in the same way as eigenvector centrality. Other centralities entail more difficulties; for example, once one decides how to transform from edge weights to edge costs (i.e., edge distances), it becomes straightforward to generalize betweenness centrality to weighted networks (because one now knows how to calculate distances), but deciding what function to use (e.g., inverting the weights or doing something else) to obtain distances in the first place involves an arbitrary decision that can severely impact the interpretation of betweenness centrality values.

In a friendship network, one may be interested in the number of people with whom an individual says they are friends (i.e., out-degree); the number of people who say that they are friends with an individual (i.e., in-degree); any type of edge, regardless of the direction; or only edges that are mutually reported (i.e., "reciprocal"). Any of these choices can be useful, depending on the question of interest, and it is important to select measures that are appropriate to one's question and context. For instance, if we seek to identify the most popular people in a school, it may be relevant to use in-degree. One way to quantify popularity is by calculating (unweighted) in-degree (e.g., by counting the number of people who say that they like the individual using a binary survey question or by thresholding a continuous "liking" rating to create an unweighted edge) or through weighted in-degree (i.e., "in-strength") centrality (e.g., by summing continuous liking ratings that an individual receives from different people; Zerubavel et al., 2015). If we are interested in understanding the spread of sexually transmitted diseases, however, we may not care about the direction of ties and opt instead to calculate degree using undirected, unweighted edges (based, e.g., on the number of sexual partners of an individual, counting any edge between two actors; Christley et al., 2005). However, incorporating directions and/or weights can become complicated for various centrality measures (both mathematically and with respect to the interpretation of centrality values), and a detailed review is beyond the scope of our paper[6].

**Recent Examples.** Recent research that examined centralities has further advanced the understanding of individual cognition in rich social environments. For instance, individuals appear to spontaneously encode and track network features of others, including eigenvector centrality (Parkinson et al., 2017), brokerage (Parkinson et al., 2017), and weighted in-degree (Zerubavel et al., 2015). Furthermore, O'Donnell et al. (2017) reported that individual differences in betweenness centrality are associated with individual differences in recruitment of brain regions during social influence (O'Donnell et al., 2017). Work on nonhuman primates illustrates that having a larger degree (which, in this study, is assignment to live in a larger group in a research colony) causally increases gray matter and resting-state functional connectivity in brain regions that are important for social functioning (Sallet et al., 2011). Although these examples highlight ways in which network analysis can advance understanding of individual cognition, it is necessary to be cautious when drawing broad inferences across such studies, given the heterogeneity of studies in design and specific choices when calculating network measures. Even the same (or similar) network measure can represent different phenomena, depending on the context of a study. For example, degree encoded the potential number of social contacts (i.e., the number of individuals who were assigned to live in the same group in a research colony, irrespective of individuals' preferences for or interactions with one another) in Sallet et al. (2011), but it encoded how much a person is liked in Zerubavel et al. (2015). Additionally, the former paper calculated undirected, unweighted degrees, whereas the latter calculated directed, weighted degrees. In many situations, results that use different centrality measures—even ones that may seem very different from each other—are likely picking up some shared information. Researchers should carefully consider these and other factors when aggregating findings across studies and forming hypotheses for future studies.

---

[6] See Wang, Hernandez, & Van Mieghem (2008) and White & Borgatti (1994) for helpful discussions.

**Community Structure and Other Large-Scale Network Structures**

Given a network, it is often insightful to study its large-scale structural patterns. Consider your own social network of friends. How might you organize the individuals in your social network? One intuitive way is to categorize your friends into groups, such as friends from high school, teammates from a sports league, fellow cosplayers, and so on. Similarly, many researchers are interested in understanding how nodes in a social network congregate into groups (Porter et al., 2009). They are also often interested in other large-scale patterns, such as core versus peripheral groups (Csermely et al., 2013; Rombach et al., 2017), the roles and positions of individuals in a network (Rossi & Ahmed, 2015; Wasserman & Faust, 1994), and so on. In the present section, we focus on the idea of algorithmically detecting tightly-knit sets of nodes called "communities"[7].

The best-studied type of large-scale structure in a network is "community structure", in which (in idealized form) densely-connected sets of nodes are connected sparsely to other densely-connected sets of nodes (Newman, 2018; Porter et al., 2009). Observing the clustered structure of a network of a school can provide insight into the features by which people organize into friendship groups (e.g., based on mutual interests or academic subdisciplines) (Traud et al., 2012). Furthermore, in a large network, finding dense communities of nodes in an algorithmic way may allow one to break down the network into smaller, manageable subsets. However, how do we identify sets of nodes that form a community in a network? There are numerous methods to detect communities in networks, including both sociocentric (i.e., complete-network) and egocentric approaches. Although the notion of communities (and related notions, such as cohesive groups; Wasserman & Faust, 1994) in a network is intuitively appealing, it is very challenging to precisely define what it means for a group of nodes (i.e., a "community") to be "densely connected" and "sparsely connected" (Fortunato & Hric, 2016). One common approach to detect communities is modularity maximization, in which one seeks a partition of a network that maximizes "modularity", an objective function that quantifies the extent to which nodes in a community connect with one another in comparison to some baseline (Newman, 2006). Another popular approach is statistical inference of communities (and other large-scale network structures) using stochastic block models (Fortunato & Hric, 2016; Peixoto, 2017). There are numerous other algorithms to identify communities (with new ones published frequently), but a review of these methods is beyond the scope of our paper[8].

---

[7] For a brief discussion of other large-scale network structures, see our Supplementary Material.

**Multilayer Networks**

Thus far, we have discussed single-layer (i.e., "monolayer") networks, as we have concentrated on networks with a single type of node in which the nodes are connected to each other with a single type of tie. Mathematically, one represents a monolayer network as a graph (Newman, 2018). However, real networks are typically more complicated, as they typically include multiple types of relationships (sometimes between multiple types of nodes) and interactions that change over time. Multilayer network analysis allows the study of richer network representations to further explore how different elements that comprise the social world interact with each another. A multilayer network consists of a set of layers that each have their own network of nodes and edges, along with interlayer edges that connect nodes from different layers[9].

As we indicated previously, individuals (i.e., nodes) in a social network can have many different types of relationships (i.e., edges). For instance, nodes that encode the individuals in a closed network (e.g., a town) can be connected to each other with edges that represent different types of relationships, such as friendship, professional ties, and recreational relationships. One can simultaneously encode all of these relationships in a multilayer network, with each type of relationship in a different layer. In our town example, each layer includes the same nodes (e.g., every townsperson), although this need not be true in general, but different layers have different types of edges (e.g., with layers 1, 2, and 3 encoding friendships, professional ties, and recreational relationships, respectively; see Figure 4). We also suppose that all interlayer edges in this example are between instantiations of the same individual in different layers. This type of multilayer network, in which different layers encode different types of relationships and interlayer connections exist only between the corresponding node across layers, is called a "multiplex" network.

---

[8] See Porter et al. (2009) for a friendly introduction to community structure and Fortunato & Hric (2016) for a recent review.
[9] For a detailed review about multilayer networks, see Kivela et al. (2014). For a recent survey, see Aleta and Moreno (2019). For a review of multilayer networks in the context of animal behavior, see Finn et al. (2019).

Multilayer networks can include different types of nodes and/or different types of nodes in different layers. Consider the online social networks of an individual. An individual may use Facebook to connect with friends but LinkedIn for professional ties. If we encode connections in these social media in a multilayer network, with the individual's Facebook and LinkedIn networks in different layers, respectively, different nodes exist in each layer and some edges may cross layers (e.g., nodes that communicate across the two platforms). Multilayer networks can also encode more complicated types of interactions. For instance, one layer may consist of friendships, with nodes encoding people and edges encoding friendships, and a second layer may consist of a network of restaurants in town, with nodes encoding restaurants and edges encoding culinary collaborations (see Figure 4). Edges between the two layers can represent which restaurants are visited by which individuals, allowing one to examine phenomena such as relationships between friendship groups and restaurant-patronage patterns.

*Temporal Networks*. In a network, nodes and edges (and edge weights) often change over time. For instance, in the social network of a town, people move in and out (changes in nodes), so the relationships between people change (i.e., time-dependent edges) over time. It is often convenient to represent a temporal network using a multilayer network, with each layer encoding the network at a specific time or aggregated over a specific time period. Research on multilayer representations of temporal networks is related to analysis of temporal networks more generally (for reviews, see Holme & Saramäki, 2012; Holme, 2015), and investigating a temporal network may be useful for researchers who seek to relate individual cognition to dynamically changing social environments.

*Learning from other fields*. As we discussed in this section, there is great potential for using multilayer networks to advance the study of complex human behavior and social systems. It seems especially promising for social neuroscientists who are interested in studying individual cognition in the context of broader social contexts. Multilayer networks can provide integrated representations of the diversity of networks that surround an individual, enabling researchers to draw insight and test how different layers of a network influence both each another and processes that occur on them. Although the analysis of multilayer networks is a relatively novel methodology in network science, it has enriched the study of diverse topics, including transportation systems (Gallotti & Barthelemy, 2015), coauthorship networks (Berlingerio et al., 2013), ecological networks (Pilosof et al., 2017), brain networks (Vaiana & Muldoon, 2018), and animal social networks (Barrett et al., 2012). Researchers who study human behavior can learn and draw inspiration from such prior work. For example, see Finn et al. (2019) for a detailed discussion of the use of multilayer network analysis to study animal behavior and Aleta & Moreno (2019) and Kivela et al. (2014) for broader reviews of multilayer networks.

## Methods to Obtain Networks

In this section, we discuss some of the most common methods for obtaining networks.

**Self-report surveys and questionnaires**. A particularly common approach for obtaining social networks is through self-report surveys and questionnaires. Using a name generator, one asks participants to list people with whom they are connected in a social network. In the same survey, one can generate multilayer networks by asking a selection of questions (e.g., "With whom are you friends?" and "To whom do you turn for advice?")[10]. Name generators can be either fixed choice (e.g., "Name the 7 people with whom you are closest.") or free choice, which does not impose limits on the number of people that a person can list. When it is possible obtain all of the names of individuals in a network prior to data collection, one can use roster-based methods. In a roster-based approach, one gives participants a list of all individuals in a network and asks them to characterize their relationship with each individual (e.g., indicating whether they are friends with each person, the strength of their friendship, and so on). Roster-based approaches have fewer recall issues than other approaches, and it is preferable to use them when possible. As with all self-reported data, all of these methods have potential concerns about bias and inaccuracy because of desirability concerns of participants and question-order effects (Pustejovsky & Spillane, 2009). However, this potential disadvantage of self-report surveys is potentially an object of interest in itself. For instance, a researcher who is interested in understanding how people understand and represent their own social networks, even if they are not accurate, can use the framework of cognitive social structures (Krackhardt, 1987).

---

[10] One should carefully consider the phrasing and ordering of questions in name generators, as these features can affect participants' responses. For detailed treatments of these issues, see K. E. Campbell & Lee (1991), Marin & Hampton (2007), and Pustejovsky & Spillane (2009).

**Direct observation**. Another method to obtain networks is through direct observation. This is a common option for researchers who study animal social networks, as they use it for observing and recording animal behavior (Noonan et al., 2014; Sallet et al., 2011), although many recent studies of animal social networks have employed technology such as radio frequency identification (RFID) data (Bonter & Bridge, 2011; Firth et al., 2017; Krause et al., 2013). In humans, direct observation can be labor-intensive and is typically feasible only for small groups. For instance, a researcher may observe the classroom behavior of children to construct a friendship network (Gest et al., 2003; Santos et al., 2015).

**Archival and third-party records.** It is also possible to reconstruct social networks using archival or third-party records. A researcher who is interested in understanding intermarriage of royal families in Europe during the 1500s can look at historical marriage records to reconstruct such a network. For instance, Padget & Ansell (1993) used historical data to characterize and analyze the social network of political elite families in 13[th] Century Florence, and they were able to identify network characteristics that contributed to the rise of the powerful Medici family. One can also leverage technological advances to obtain data such as e-mail, phone, and geographic-location records to reconstruct not only networks that encode the existence of communication ties, but also the frequency and patterns of communication. This approach has been used for studying communication within organizations (C. S. Campbell et al., 2003), face-to-face contact in academic conferences and museums (Isella et al., 2011), and features of social structures that are inferred from mobile-phone data (Eagle & Pentland, 2006).

Advantages of these methods include that they do not rely on self-reporting, are relatively low-effort (although such data may be hard to access), and can provide a wealth of different types of data (and an abundance of data of each type). However, researchers should be mindful when interpreting the social significance of a tie in networks that they construct using these approaches. For instance, an e-mail exchange in an organization may encode only formal ties between coworkers and fail to capture less formal ties, which can also affect the phenomena that a researcher is hoping to capture. Perhaps an employee exchanges frequent e-mails with their supervisor and none at all with a coworker (with whom they may have a closer relationship) who sits in the cubicle next to them. Consequently, measuring the distance between people in a network that one constructs using exclusively e-mail data is unlikely to provide a complete picture of these individuals' social relationships. Therefore, researchers should be mindful of these considerations when drawing inferences from calculations that use such networks. Researchers should also be mindful of privacy concerns that may arise from accessing potentially sensitive personal information of participants, particularly when considering posting data online (which ordinarily is desirable, as it helps promote open science initiatives). It is possible to reconstruct even fully anonymized data, especially when there is a lot of data for each participant, to identify individuals (Herschel & Miori, 2017).

The rise of online social networking websites, such as Facebook and Twitter, has also afforded researchers the opportunity to "scrape" them (and otherwise acquire data from them) and study online social networks (Lewis et al., 2008), although the policies of the companies that own the networks may entail some limitations. Additionally, when studying large online social networks, it is also necessary to pay close attention to the characteristics both of the network at large and of smaller local networks of interest, as these they may influence salient network measures (see, e.g., Jeub et al., 2015; Ugander, Backstrom, Marlow, & Kleinberg, 2012). Furthermore, social networks obtained from online websites are often 1-ego networks[2] (encoding information about an individual ego and their friends), which have limitations, as discussed in our "Sociocentric Networks versus Egocentric Networks" section. One also needs to be careful when interpreting the social significance of ties in online social networks. For instance, a large degree on Facebook or Twitter may be an indication that an individual frequently uses the platform, rather than being related to the types of individual differences in socio-behavioral tendencies that may be of more interest to social neuroscientists. For example, a person with a small degree (i.e., few "friends") on Facebook may actually have a large degree in their offline life. This can be problematic if one uses degree from Facebook data alone as a measure to relate to a neural or behavioral measure. More generally, there can be additional uncertainty in effects that one infers from data from social networking websites, because such effects only characterize a small slice of individuals' social worlds (Ugander et al., 2012). Although this issue is particularly salient for nuances of analyzing online-social-network data, researchers need to be careful more generally to ensure that they are obtaining sufficient relevant information about an individual's social world whenever they attempt to relate individual differences in network centrality values (or other differences in individuals' network characteristics) to neural data or socio-behavioral tendencies. Similar issues can arise if one uses individual differences in centrality measures (e.g., degree) based on a bounded social group (e.g., a school), while failing to capture sufficiently many relevant aspects of individuals' social worlds. For example, in an analogous offline situation to the aforementioned online one, an individual may have small degree in their school but have many friends outside of school who are not captured if one calculates degree based only on a school network. Therefore, when researchers are interested in interpreting a difference in social network position[3] as an individual difference measure (i.e., a trait), it is advantageous to construct network data that captures people's full social worlds. When this is not possible (as is often the case), it is desirable to ask participants about their relationships outside of the social network that one is obtaining.

## Tutorial: Example Social Network

Now that we have discussed some key concepts in network analysis that are particularly relevant for people who are interested in studying human social networks, we present a tutorial using a sample network. In this artificial network, we are interested in characterizing the network of a dorm (with 50 students). Suppose that we obtained this data by asking participants to go through the list of everyone in the network and identify whether they are friends with each individual (i.e., using a roster-based approach). This gives directed edges, because some friendships may not be reciprocated. If we are interested in understanding how individuals cognitively represent different members of the network or how individual differences in network measures are correlated with differences in neural or behavioral variables, we can also obtain brain data from all or some of the network members. (We do not cover this idea in the tutorial.) The tutorial uses an artificial network with 50 nodes, which we label with people's names to facilitate exposition. We use the IGRAPH package in R (Csardi & Nepusz, 2006) to visualize the data and calculate various network measures—such as degree, eigenvector centrality, and betweenness centrality—and to illustrate community detection. Our tutorial includes detailed comments on the practical application of the concepts that we have discussed in this paper. We also present a separate tutorial to illustrate visualization of multilayer networks using the PYMNET library in Python (Kivelä, 2017). Both tutorials are available at https://github.com/elisabaek/social_network_analysis_tutorial. We hope that they will be helpful for researchers who are interested in incorporating network measures in studies of individual cognition.

## Future Directions

In the present paper, we have given an introductory overview of basic network ideas and concepts that we hope will provide a helpful starting point for social neuroscientists who are new to network analysis. Although the incorporation of network-analysis tools in social neuroscience is in its nascent stages, recent work using such tools has produced fascinating insights into how features of an individual's social world are reflected in their brain. There are many open questions in the area, so it is a particularly exciting time to do research in it. In this section, we highlight areas for future growth. We discuss both how social neuroscientists can integrate common network methods in new lines of inquiry and how to productively incorporate new developments and tools in network science and mathematics into future work in social neuroscience.

**Open questions that leverage existing network tools**. We begin by highlighting some of the many open questions in social neuroscience that can benefit from network analysis. Although we will of course not be exhaustive, we hope to highlight the broad range of exciting research opportunities for social neuroscientists who are interested in using network analysis.

*Information about different types of relationships.* Several of the findings that we discussed highlight how the brain has mechanisms to track and spontaneously retrieve information about different aspects of friendship networks, such as the extent to which individual members are popular (Zerubavel et al., 2015), socially valuable (Morelli et al., 2018), well-connected to well-connected others (Parkinson et al., 2017), and serve as brokers (Parkinson et al., 2017). These studies barely scratch the surface of the many different types of information about the social world that our brains may track. People's lives consist not only of different types of social groups (e.g., friendship, professional, and family), but also different types of information about the same social groups that may be important for successful social navigation. For instance, in the same group of friends, individuals may turn to different people when seeking emotional support versus career advice. Indeed, recent findings suggest that centralities in a social network can have different implications, depending on how one characterizes relationships. For example, Morelli et al. (2018) examined in-degree in two different social networks—one with edges that encode trust and the other with edges that encode shared fun—in the same college dorms. People with better well-being were located more centrally in the fun network, and people with higher empathy were located more centrally in the trust network. Such findings suggest that where an individual is located in different social networks (i.e., with different types of edges) of the same social group is associated with different behavioral outcomes. Although this was not tested by Morelli et al. (2018), one possibility is that perceivers also track the centralities of others in the different networks (e.g., those with trust relationships versus those with fun relationships), as this information may be important for guiding behavior in different contexts. For example, when seeking empathic support, it seems advantageous to seek individuals who are central in a trust network. However, when looking to have fun, one might seek individuals who are central in a fun network. It may be particularly fruitful to conduct studies that explore how individual brains encode and retrieve information about social networks with different types of connections in the same social group.

Characterizing different types of relationships in a social group may also improve understanding of not only who is popular, but also those to whom others turn for support or empathy. Given that individuals who are more likely to seek social support to help regulate their emotions (i.e., interpersonal emotion regulation) tend to have better well-being and more supportive relationships (Williams et al., 2018), one fruitful future direction may be to use centrality measures to identify supportive individuals (see, e.g., Morelli et al., 2018) and test how people's cognitive and affective processes are affected by their social distance to these individuals or by the amount of time spent with these individuals (e.g., by incorporating weighted edges).

*Individual differences in network features.* A small body of research has also begun to explore associations between individual differences in network positions and brain activity. Popular individuals (specifically, individuals with large in-degree in a network in which edges represent being liked by others) tend to have greater neural sensitivity in the brain's valuation system in tracking the popularity of others in a network (Zerubavel et al., 2015), and people with higher brokerage (which they examined by calculating an egocentric betweenness centrality in a Facebook friendship network) exhibit greater activity in the brain's mentalizing system when considering and incorporating social recommendations to make their own recommendations of consumer products to others (O'Donnell et al., 2017). It has also been illustrated that social status in non-human primates covaries with structural and functional differences in brain regions that are associated with social cognition (Noonan et al., 2014). In combination, these findings suggest that an individual's social-network position is associated with neural and behavioral responses to various everyday situations. There are many open questions, as only a few studies have related individual differences in social-network position to neural responses, and even fewer have done so in the context of social decision-making. Future studies that explore how individual differences in social-network position relate to neural responses during social tasks (e.g., social influence, emotion regulation, and interpersonal communication) may be particularly fruitful. Findings from such studies have the potential to advance understanding of how particularly influential individuals may be distinctive in how they use their brains and in their responses to various social situations.

*Causal relationships*. Most research that integrates neuroscience with social-network analysis has been cross-sectional (see Table 2). Accordingly, there remain many questions about the causal directions of the various correlative findings that we have discussed in this paper. It remains unclear, for instance, whether differences in neural responses cause or result from differences in social-network characteristics. Experimental findings from nonhuman primates offer some clues, as it has been demonstrated that social-network characteristics (e.g., network size) causally affect the structure and functional responses in regions of the macaque brain that are associated with social cognition (Sallet et al., 2011). Although long-term, meaningful experimental manipulation of social networks in humans is very challenging to implement because of practical and ethical concerns, longitudinal studies can also elucidate some of the ambiguity about causality. Longitudinal studies that span key neural and social developmental periods, such as adolescence or older adulthood, may be particularly fruitful for providing insight into questions about the causal directions of effects.

Despite the challenging nature of experimental manipulation of social networks in humans, there are a few possible approaches to pursue. One possibility is to recruit participants to join either offline or online interest-based communities and then randomly assign participants to different social networks that one controls experimentally to vary network characteristics of interest. For example, perhaps one wants a network to have a specific degree distribution, such as many people with small degrees and few people with large degrees. Such methods have been used previously to test how social-network characteristics influence the spread of behavior in online social networks (e.g., how similarity of contacts influence adoption of health behavior; Centola, 2010, 2011), but to our knowledge they have not yet been used with neuroimaging tools. Future studies that use similar experimental methods while also obtaining neural responses before and after individuals' experiences in a social network may further elucidate the causal directions of such observations. However, it remains unclear whether (and to what extent) an individual's cognitive and affective processes are influenced by artificially constructed social networks. Nevertheless, if successful, future studies that employ such approaches may provide valuable insights into causal relationships between social and neural phenomena.

**Potential of incorporating new methods of network analysis**. We now briefly overview a few new methods in network analysis and related subjects that may be insightful for developing richer characterizations of social-network structures. We keep our descriptions brief because of the introductory nature of this paper.

As we discussed in previous sections, multilayer and temporal networks afford rich opportunities to examine how individual brains interact over time with the social world in which they live. For instance, multilayer network analysis will be useful for longitudinal studies to help understand how characteristics of a social network change over time, so such analysis may be able to inform causal relationships that characterize some of the previous findings that link brain activity and social-network characteristics. One can potentially use multilayer networks to examine interactions between brain networks and social networks over time to help predict behavior. It is also possible to analyze cognitive social structures using multilayer networks (Kivelä et al., 2014). Tools from network science (including multilayer network analysis) have been used to analyze functional and anatomical networks in the brain (Bassett et al., 2011; Fair et al., 2008; Hutchison et al., 2013; Vaiana & Muldoon, 2018; van den Heuvel & Sporns, 2013), as well as to link these brain networks with social-network structures (Schmälzle et al., 2017) and with cognition and behavior (Bassett & Mattar, 2017; Mattar et al., 2018). Recently, researchers have highlighted potential benefits of using multilayer network analysis to represent such complex relationships, and these efforts have the potential to advance understanding of processes of interest to social neuroscientists (Falk & Bassett, 2017). One potential fruitful application is investigating how health behaviors change over time (Christakis & Fowler, 2007). For instance, one can use multilayer and temporal networks to study how to predict health-behavior change (e.g., quitting smoking) from changes in an individual's social network (e.g., joining a support group to stop smoking) through changes in functional networks in the brain (e.g., how regions in the brain's valuation system respond to smoking cues). Such a research question can contribute to broader understanding of how people's social environments impact neural processing and behavior.

For a brief discussion of additional network-analysis approaches—such as using hypergraphs, topological data analysis, community-level characteristics, and other mesoscale features—that may be fruitful for characterizing social networks for social neuroscience applications, see our Supplementary Material.

**Conclusions and Outlook**

Recent research in social neuroscience that relates characteristics of people's social networks to their individual cognition offers new insights into how the brain represents and may be influenced by its social context. Tools from network analysis provide rich opportunities for social neuroscientists who are interested in (1) studying how people navigate and interact with their complex social environments and (2) the mental architecture that supports these processes. Researchers can leverage existing and developing tools and measures in network analysis to study new questions. Findings from such studies can contribute to relevant theories in numerous areas in psychology, neuroscience, and related fields. For instance, insights from network analysis can inform theories of individual cognition, interpersonal relationships, and social influence (e.g., through relating features of individuals' social worlds to how they use their brain in certain contexts, through observing how social network distance influences how people process the world, and through understanding how people in specific network positions use their brains differently). The use of network analysis in social neuroscience is in its emerging stages, so this is a particularly exciting time, with many opportunities to contribute to shaping the direction of the field.

SOCIAL NETWORK ANALYSIS FOR SOCIAL NEUROSCIENTISTS                                      35Campbell, K. E., & Lee, B. A. (1991). Name generators in surveys of personal networks. *Social Networks*, *13*(3), 203–221. https://doi.org/10.1016/0378-8733(91)90006-F

Centola, D. (2010). The spread of behavior in an online social network experiment. *Science*, *329*(5996), 1194–1197. https://doi.org/10.1126/science.1185231

Centola, D. (2011). An experimental study of homophily in the adoption of health behavior. *Science*, *334*(6060), 1269–1272. https://doi.org/10.1126/science.1207055

Cherkassky, B. V., Goldberg, A. V., & Radzik, T. (1996). Shortest paths algorithms: Theory and experimental evaluation. *Mathematical Programming, Series B*, *73*(2), 129–174. https://doi.org/10.1007/BF02592101

Christakis, N. A., & Fowler, J. H. (2007). The spread of obesity in a large social network over 32 years. *The New England Journal of Medicine*, *357*(4), 370–379. https://doi.org/10.1056/NEJMsa066082

Christley, R. M., Pinchbeck, G. L., Bowers, R. G., Clancy, D., French, N. P., Bennett, R., & Turner, J. (2005). Infection in social networks: Using network analysis to identify high-risk individuals. *American Journal of Epidemiology*, *162*(10), 1024–1031. https://doi.org/10.1093/aje/kwi308

Coviello, L., Sohn, Y., Kramer, A. D. I., Marlow, C., Franceschetti, M., Christakis, N. A., & Fowler, J. H. (2014). Detecting emotional contagion in massive social networks. *PLoS ONE*, *9*(3), e90315. https://doi.org/10.1371/journal.pone.0090315

Crossley, N., Bellotti, E., Edwards, G., Everett, M. G., Koskinen, J., & Tranmer, M. (2015). *Social network analysis for ego-nets*. SAGE Publications.

Csardi, G., & Nepusz, T. (2006). The igraph software package for complex network research. *InterJournal, Complex Sy*, 1695. http://igraph.org

Csermely, P., London, A., Wu, L. Y., & Uzzi, B. (2013). Structure and dynamics of core/periphery networks. *Journal of Complex Networks*, *1*(2), 93–123. https://doi.org/10.1093/comnet/cnt016

SOCIAL NETWORK ANALYSIS FOR SOCIAL NEUROSCIENTISTS    37others. *Proceedings of the Royal Society B: Biological Sciences*, *284*(1854), 20170299. https://doi.org/10.1098/rspb.2017.0299

Fortunato, S., & Hric, D. (2016). Community detection in networks: A user guide. *Physics Reports*, *659*, 1–44. https://doi.org/10.1016/j.physrep.2016.09.002

Fowler, J. H., & Christakis, N. A. (2008). Dynamic spread of happiness in a large social network: Longitudinal analysis over 20 years in the Framingham Heart Study. *BMJ*, *338*(7685), 23–26. https://doi.org/10.1136/bmj.a2338

Gallotti, R., & Barthelemy, M. (2015). Anatomy and efficiency of urban multimodal mobility. *Scientific Reports*, *4*(1), 6911. https://doi.org/10.1038/srep06911

Gest, S. D., Farmer, T. W., Carolina, N., & Cairns, B. D. (2003). Identifying children's peer social networks in school classrooms: Links between peer reports and observed interactions. *Social Development*, *12*(4), 513–529. https://doi.org/10.1111/1467-9507.00246

Grund, T. U. (2012). Network structure and team performance: The case of English Premier League soccer teams. *Social Networks*, *34*, 682–690. https://doi.org/10.1016/j.socnet.2012.08.004

Hampton, W. H., Unger, A., Von Der Heide, R. J., & Olson, I. R. (2016). Neural connections foster social connections: A diffusion-weighted imaging study of social networks. *Social Cognitive and Affective Neuroscience*, *11*(5), 721–727. https://doi.org/10.1093/scan/nsv153

Herschel, R., & Miori, V. M. (2017). Ethics & big data. *Technology in Society*, *49*, 31–36. https://doi.org/10.1016/j.techsoc.2017.03.003

Holme, P. (2015). Modern temporal network theory: A colloquium. *The European Physical Journal B*, *88*(9), 234. https://doi.org/10.1140/epjb/e2015-60657-4

Holme, P., & Saramäki, J. (2012). Temporal networks. *Physics Reports*, *519*(3), 97–125. https://doi.org/10.1016/j.physrep.2012.03.001

House, J. S., Landis, K. R., & Umberson, D. (1988). Social relationships and health. *Science*, *241*(4865), 540–545.

SOCIAL NETWORK ANALYSIS FOR SOCIAL NEUROSCIENTISTS 43

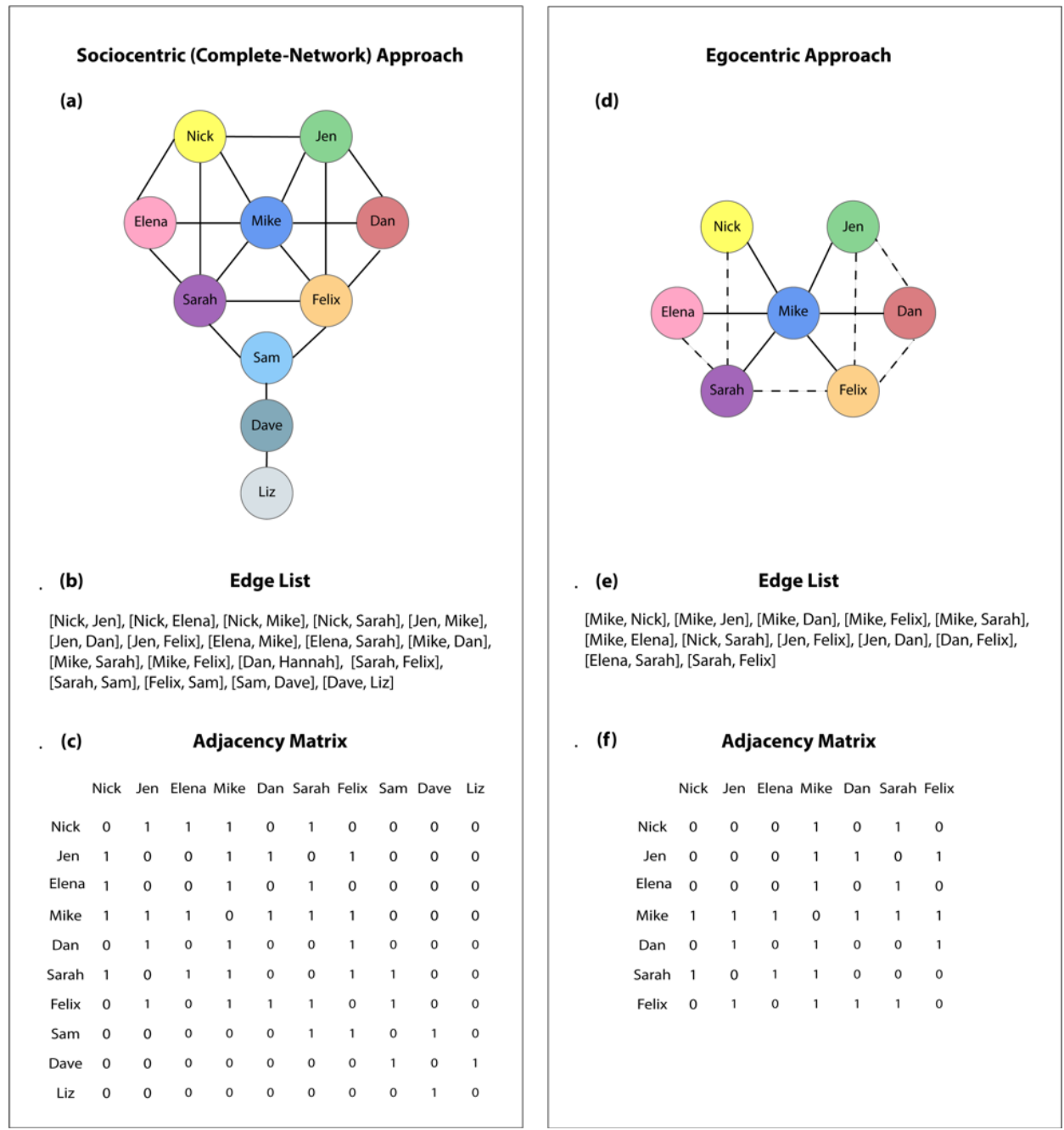

**Figure 1.** Approaches to study and mathematically represent social networks. **(a–c)** In a sociocentric approach, one characterizes relationships between all members of a bounded social network. **(a)** A graphical representation of an undirected, unweighted sociocentric network that represents friendships between members of a bounded community. The colored circles are nodes (also called vertices), which represent individuals in the social network. The lines between the nodes are edges, which represent friendships or some other relationship between individuals. **(b)** One can also represent networks with an edge list, which is a list of all direct connections between nodes. **(c)** It is also common to represent an *n*-node network with an adjacency matrix **A** of size *n* x *n* (with *n* = 10 in this example). The elements $A_{ij}$ of **A** encode the edges (both their existence and their weights) between each node pair (*i,j*) in a network. In an undirected, unweighted network (such as the depicted one), an associated adjacency matrix is symmetric. For example, the edge between Nick and Jen yields a 1 in the associated element of an adjacency matrix. **(d–f)** In an ego-network approach, one characterizes relationships in a network from an ego's point of view. Suppose that we obtain information about the same social network as the one in the left column from interviewing only Mike, a single member of the network. This gives us Mike's ego network. We draw solid lines to represent Mike's responses about his direct friendships and dotted lines from Mike's responses about whether his friends are also friends with one another. Comparing the graph from the sociocentric and ego-network approaches illustrates that the latter is missing information about several existing edges between nodes (e.g., between Nick and Elena, Nick and Jen, and so on). We also see this in the ego network's associated **(e)** edge list and **(f)** adjacency matrix.

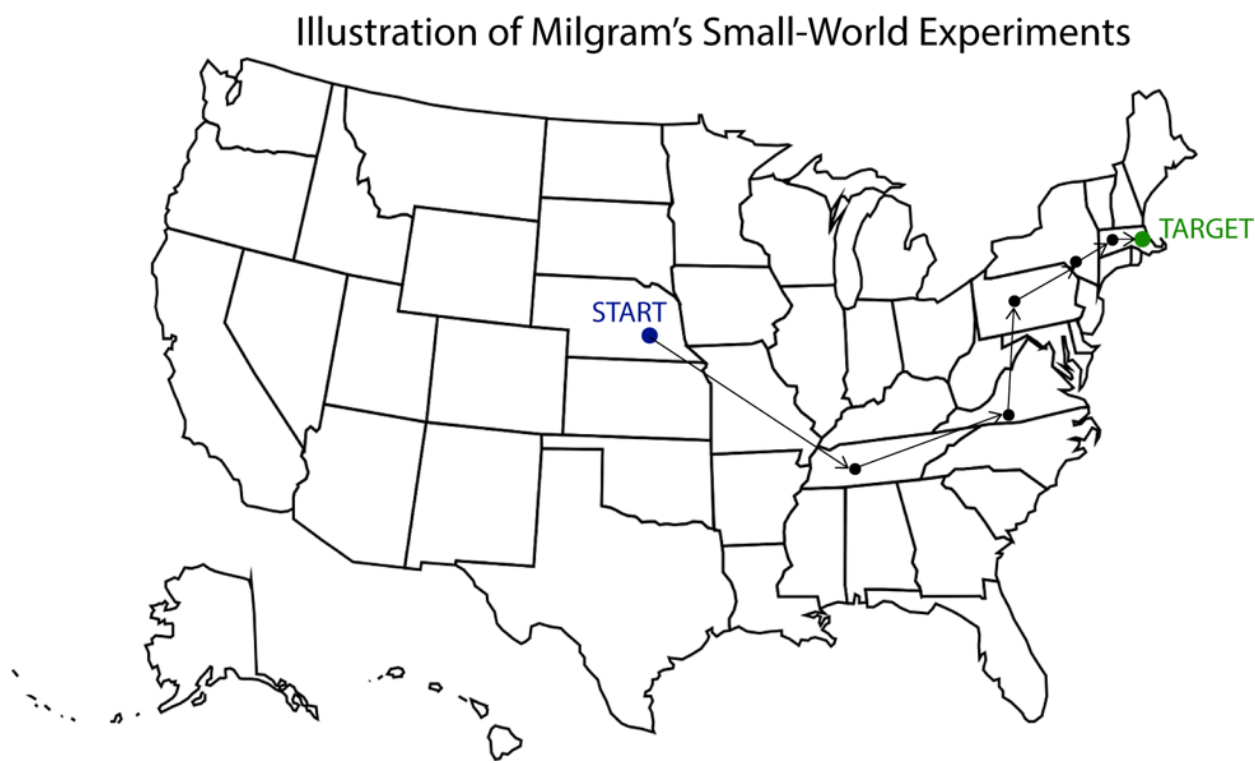

**Figure 2. An illustration of Stanley Milgram's Small-World Experiments that Demonstrate Social Distance.** In their pioneering studies of social distance, social psychologist Stanley Milgram and colleagues (1967,1969) concluded that, on average, people are separated by six or fewer social connections. As our illustration demonstrates, individuals in the midwestern United States (the starting position) were able to send a package to a stranger in Massachusetts (the target individual) through a path whose length was about 6. In one experiment, of the 160 packages that started in Nebraska (the starting position in this figure), 44 packages successfully arrived at the target individual. Of these 44 packages, the mean number of edges was about 6. Milgram's small-world experiments illustrate unweighted social distance in a real-life context.

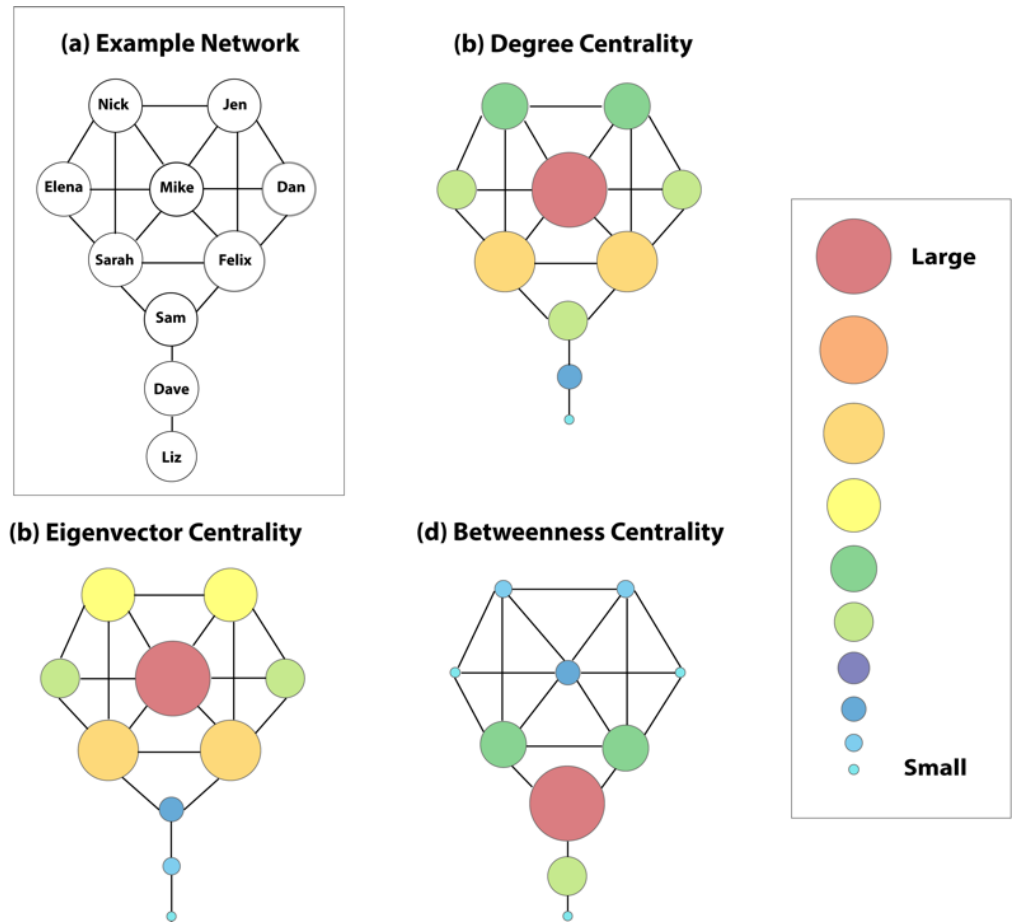

**Figure 3.** A few common measures of centrality. This adapted version of Krackhardt's kite graph (Krackhardt, 1990) illustrates several variants of centrality. **(a)** An example friendship network, with each node labeled with the name of an individual. **(b–d)** Variations of the same network, with the nodes resized to reflect the value of a particular centrality measure. **(b)** Degree centrality (i.e., degree) is the number of other nodes to which a node is connected directly (i.e., adjacent). Mike has a degree of 7, the largest value in the network. **(c)** Eigenvector centrality captures how well-connected a node is to well-connected others. Although Elena, Dan, and Sam all have the same degree (of 3), Sam has a much smaller eigenvector centrality, as his friendships are with relatively poorly connected individuals. **(d)** Betweenness centrality captures the extent to which a node lies on shortest paths between pairs of nodes. Sam has the largest betweenness centrality in this network, because he connects many nodes in the network that otherwise would be on disconnected components of the network.

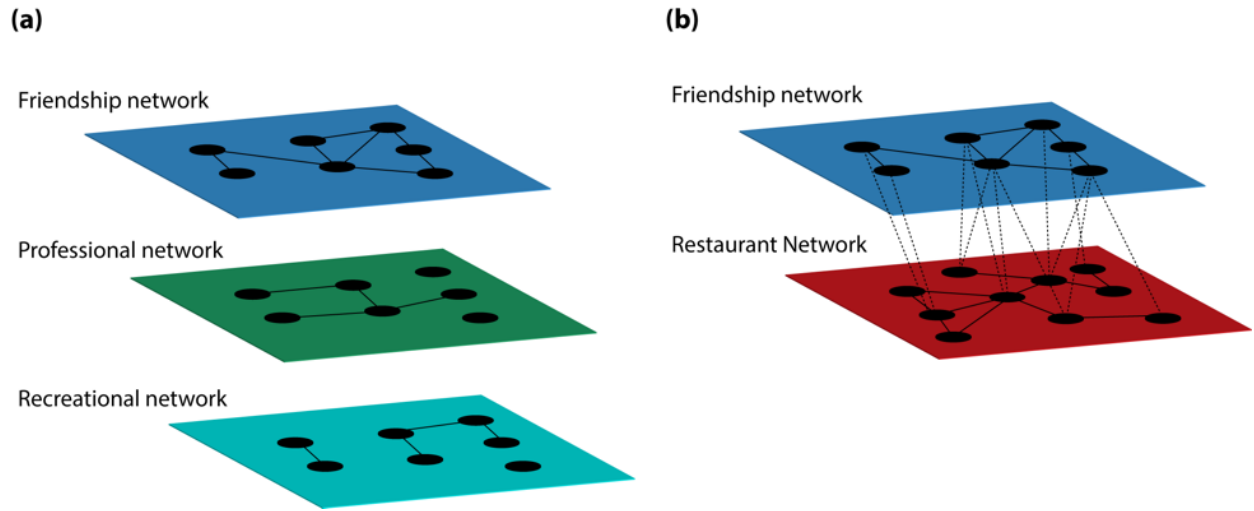

**Figure 4.** Examples of multilayer networks. **(a)** A multiplex network is a type of multilayer network in which each layer has a different type of edge and interlayer edges can exist only between corresponding nodes in different layers. The nodes in this example represent the same individuals in each layer, and the edges in different layers encode different types of social relationships. We do not show any interlayer edges. In the first layer, edges encode friendships between individuals, whereas edges encode professional relationships between individuals in the second layer and recreational relationships between individuals in the third layer. **(b)** In this more general example of a multilayer network, the first layer encodes the same friendship network that we showed in panel **(a)**. The second layer represents a restaurant network, where nodes represent restaurants and intralayer edges encode culinary collaborations between restaurants. Interlayer edges encode restaurant patronage of a restaurant by an individual, with an edge indicating that an individual has visited a restaurant. This type of multilayer network can help one understand possible relationships between friendship and restaurant-patronage patterns. In this example, friends tend to eat at the same restaurants.

**Box 1. Computing distances between people in a weighted social network: An example using Dijkstra's algorithm.**

Social neuroscientists are often interested in characterizing not just the existence of social ties between people, but also the relative strengths of those ties (i.e., in constructing a weighted social network). It is important to consider the consequences of representing a network using weighted ties for calculating and interpreting quantities like social distances between people in the network. We outline a common method for calculating distance in a weighted network using Dijkstra's shortest-path-first algorithm (Dijkstra, 1959; Newman, 2001) and consider its implications.

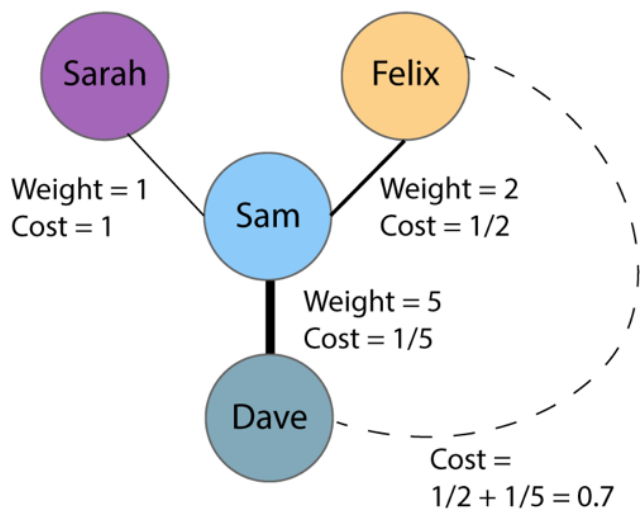

Dijkstra's algorithm works by finding a path of "least resistance" between two nodes, where the "resistance" is the cost of traversing a path between two nodes. In a weighted network, the simplest choice for the cost of a tie between two nodes is the inverse of the tie's weight, where larger weights represent stronger ties and associated lower costs. For instance, given that the weight of the edge between Felix and Sam is 2 and that the weight of the edge between Sam and Dave is 5, the associated costs are 1/2 and 1/5, respectively. For indirect connections between two nodes (i.e., paths that require at least two edges), one calculates cost as the sum of the costs of the direct ties between nodes. In the example above, Felix and Dave are connected through the edge Felix → Sam of weight 2 and the edge Sam → Dave of weight 5. Therefore, the cost of the 2-step path Felix → Sam → Dave is 1/2 + 1/5 = 0.7.

In many situations, Dijkstra's algorithm may identify two nodes that are connected only indirectly as "closer" than two nodes that are connected directly. In the example above, Felix and Dave are not connected directly and have a cost (i.e., distance) of 0.7. Consider another pair of nodes, Sarah and Sam, who are connected directly with an edge of weight 1. This

> yields a cost of 1 for their edge, because 1/1 = 1. In Dijkstra's algorithm, Felix and Dave are considered to be *closer* to each other than Sarah and Sam are to each other, even though Felix and Dave are not connected directly.
>
> This is an important implication of what measures to consider, as many social neuroscientists may want to consider directly connected people (e.g., friends in a friendship network) as closer than indirectly connected people, a premise that fits well with the types of applications and research questions that are common in the field. For researchers who are interested in understanding the spread of phenomena (e.g., information or a disease) in networks, Dijkstra's algorithm may give a helpful estimate of distance because, for example, information is more likely to spread faster through edges that represent very frequent interactions than through ones that represent infrequent interactions. We encourage researchers to be driven by their research questions when making decisions about which network measures to use. We also encourage them to be attuned to the details of methods before applying and drawing inferences from them.

Table 1. Some Key Terms in Network Analysis.

| Network Term | Definition | Applications and Related Concepts |
|---|---|---|
| Network (i.e., graph) | A collection of entities (i.e., nodes) that are connected to one another (through edges). | In the context of social systems, a network consists of people (or animals) who are connected to one another. |
| Node (i.e., vertex) | A node is an entity in a network. | Most typically, a node represents a person in a social network. Nodes are also called "actors" in the context of social systems. |
| Edge (i.e., tie, link) | A connection between two entities in a network. | In a social network, an edge typically represents some type of a relationship (e.g., friendship, professional relationship, or physical encounters per day) between individuals. |
| Directed edge | A connection between two entities in a network that has an orientation. One typically uses an arrow to represent the direction of the orientation. | In the context of a social network, directed edges can be useful for characterizing concepts such as "popularity". For instance, a researcher may choose to define the popularity of an individual by the number of nominations that they receive from others in a network. |
| Undirected edge | A connection between two entities in a network that has no direction. | Edges can be undirected because the criterion that one uses to define them is undirected in nature (e.g., an edge can represent the presence of a group affiliation) or because of researcher choice (e.g., a researcher may choose to define friendship by counting only mutually reported relationships). |
| Weighted Edge | A connection between two entities in a network that encodes the strength of a relationship (or interaction). | A researcher may use subjective ratings of closeness to represent strengths of friendships in a social network. |
| Unweighted Edge | A connection between two entities in a network that does not incorporate the strength of a relationship (or interaction). | Edges can be unweighted by nature (e.g., if an edge encodes whether a relationship exists or does not exist), or by researcher choice (e.g., a researcher may choose to use an edge to represent a relationship only if it equals or exceeds a minimum threshold on the number of interactions). |
| Sociocentric Network (i.e., complete network) | Encapsulates a complete picture of who is connected to whom in a network. | An example of a sociocentric network approach is to survey all of the members of a sports team to characterize a friendship network by asking people who their friends are. |
| Egocentric Network | A network that is based on an individual ("ego") and their friends ("alters"). | An example of an egocentric-network approach is to ask one individual (the "ego") about the people (the "alters") to whom they are connected directly. In some cases, one also collects information about whether the alters themselves are connected to one another. |
| Adjacency Matrix | A mathematical representation of a network. An adjacency matrix **A** of a network is an *n* x *n* matrix (where n is the number of nodes) with elements $A_{ij}$. | See Figure 1 for examples of adjacency matrices. |
| Edge List | An edge list is a list of node pairs that are connected directly by edges. | See Figure 1 for examples of edge lists. |
| Distance | In an unweighted network, the distance between two nodes is the smallest number of edges that one needs to traverse to connect the two nodes (i.e., a shortest path). If edges are weighted, one uses associated edge costs to calculates distances. | Two nodes can be connected by direct ties (e.g., "friends", with a distance of 1) or by indirect ties (e.g., "friends of friends", with a distance of 2). Researchers should carefully consider context before drawing inferences based on distances between nodes, as interpretations of distance can be affected by various features of a network. |
| Centrality | Captures importance of actors (or of edges between them) in a social network. | There are many variants of centrality. We discuss several common types. |
| Degree Centrality (i.e., degree) | The number of edges that are attached to a node. | In a social network, an individual's degree centrality is the number of connections that they have. |
| Eigenvector Centrality | The components of the leading eigenvector of a network's adjacency matrix **A**. | Eigenvector centrality captures how well-connected an individual is to well-connected others. PageRank is an important variation of eigenvector centrality that has been used most famously to rank search results on the World Wide Web. |
| Diffusion Centrality | Captures an individual's centrality with respect to a simple spreading process on a network. | Diffusion centrality may be useful for characterizing how central individuals are in spreading items (such as information) in a dissemination process. |
| Betweenness Centrality | Measures the extent to which shortest paths between pairs of nodes traverse a node. | An individual with large betweenness centrality may have a high capacity for brokerage because more of their friends have to go through them to communicate with one another. (However, a large betweenness centrality does not necessarily entail high brokerage. See the main text for important caveats in interpreting betweenness centrality.) |
| Community | A set of nodes that are densely connected with one another but sparsely connected with other communities of nodes. | For instance, given an individual's social network, community-detection algorithms can help identify different groups of friends (e.g., friends from high school, teammates from a recreational sports league, and so on). |
| Multilayer Network | A network with multiple layers. Each layer has its own sets of nodes and edges, and there are also interlayer edges that connect nodes rom different layers. | Multilayer networks can encode social networks with many different types of relationships. For examples, see Figure 4. |

Table 2. Limitations and Challenges

| |
|---|
| The incorporation of network-analysis tools to study social systems has the potential to greatly enrich the study of the individual cognition in the context of real-life social environments. However, there are many issues for researchers to consider when making decisions about using network-analysis tools to study social systems. |
| *Challenges in data collection.*<br><br>Combining the methods that we described in our "Methods to Obtain Networks" section with neuroscientific data typically requires having collected data on neuroimaging study participants' social relations. Most existing data sets from social neuroscience studies do not have such data on participants. Consequently, it is typically necessary for a research team to acquire social network data on neuroimaging participants as part of data collection (rather than working with existing data sets). This has the potential to pose additional logistical challenges during data collection. |
| *When network tools may not be the most appropriate.*<br><br>Sometimes, it may be possible to answer a question of interest more readily by relating brain activity to other individual difference measures that may be easier to obtain than network data. For instance, if we are interested in understanding relationships between social support and brain activity, we can test the relationship between degree centrality and brain activity (inferring that smaller degree centrality entails fewer friends, which in turn entails less social support). However, it may be easier (and perhaps more appropriate, in some cases) to simply ask individuals about their subjective perceptions of social support. |
| *Causal inferences.*<br><br>As we discuss in our "Future Directions" section, researchers should be very careful when inferring (or implying) causal directions in relating brain activity and network features. Most existing studies in social neuroscience that have related brain activity and network features are cross-sectional in nature, so associated causal relationships are unclear. This arises because meaningful experimental manipulation of social network features in humans is challenging (for both practical and ethical concerns), and it can also be difficult to conduct (or otherwise obtain) longitudinal studies that involve both brain activity and social networks. |

Supplementary Material

**An Example of Other Measures of Centrality: PageRank**

As we described in the main text, there are many ways of characterizing centralities (i.e., importances) of people in a social network. We outlined a handful of such measures (specifically, degree, in-degree, out-degree, diffusion, betweenness, and eigenvector centrality) in the main text. We now describe a popular measure of centrality known as PageRank centrality (which is a variant of eigenvector centrality; Gleich, 2015). PageRank centrality (or simply "PageRank") incorporates a probability for the walking robot that we described in the "Centrality" section of the main text to "teleport" to random nodes in a network in addition to traversing the network's edges. A node (e.g., a web page) tends to be central according to PageRank if it has large in-degree (e.g., many other web pages point to it) and the incoming edges are from nodes that themselves have a large in-degree (e.g., the web pages that point to it have a lot of other web pages that point to them). PageRank takes into consideration both the direction and the weights of edges, and one construes a web page to be important if many other important web pages link to it. Suppose that a robot is randomly surfing the web, so it is randomly walking from one node (i.e., web page) to another through directed edges (i.e., hyperlinks that point from one web page to another) and randomly "teleporting" to other web pages by opening a new browser window. One can calculate the PageRank centrality of a web page by examining how often the robot visits the web page, including through teleportation, if it surfs the web forever (Masuda et al., 2017). PageRank is associated most famously with ranking web pages, but it has also been applied to investigate questions in a large variety of different fields, including ranking the influence of Twitter users, ranking academic journals and doctoral programs, and finding correlated genes and proteins. For a review of PageRank, see Gleich (2015).

**Additional Future Directions: Other Methods of Network Analysis**

We briefly discuss other methods (including ones under rapid development) in network analysis that may be useful for social neuroscientists who are interested in characterizing real-world social networks and relating those characteristics to neuroscientific data. We discuss the potential utility of tools from topological data analysis, community-level characteristics, and other mesoscale features to study social networks.

*Beyond pairwise connectivity in networks.* We anticipate that it will also be fruitful to examine relationships between nodes beyond the usual pairwise connections. The simplest way to do this is with hypergraphs, which allow edges (which are called "hyperedges" in this context) to connect more than two nodes and are thus useful for representing relationships that involve more than two people (Newman, 2018). One example is a coauthorship network, where a single hyperedge connects all of the coauthors of a manuscript, instead of connecting them through multiple pairwise edges. Another example is a network of college roommates, where it may be desirable to use a single hyperedge to connect all occupants of one room, which may be shared by more than two people. Moreover, it is possible that some pairs of roommates may also be connected directly in a pairwise fashion, so using hypergraphs gives a sensible way to simultaneously include both pairwise connections and other connections in a network structure. A more complicated, but likely very useful, approach to study relationships among arbitrarily many actors in a social network is to use "simplicial complexes" (Ghrist, 2014), an idea from algebraic topology that many researchers have leveraged for "topological data analysis" (Otter et al., 2017; Topaz, 2016). One can use tools from topological data analysis to systematically examine a diversity of structural features of social networks, such as by algorithmically finding topological "holes" (e.g., gaps) in coauthorship network (Carstens & Horadam, 2013). Perhaps such holes may help uncover barriers to academic collaboration, and it seems plausible to try to relate such topological holes to Burt's notion of "structural holes" in social networks (Burt, 1992). The most popular approach from topological data analysis is "persistent homology", which allows one to track many types of topological holes over multiple scales in a network. We anticipate that persistent homology and other tools from topological data analysis will be used increasingly in the study of social networks. See Topaz (2016) for a brief popular introduction and Otter et al. (2017) for a more mathematical introduction and a tutorial of available software.

*Community-level characteristics and mesoscale network structures.* As we discussed briefly in the main text, one can examine densely connected communities of nodes in a network. There are numerous algorithms to study community structure; some of them involve assigning nodes to single communities, and others allow overlapping communities (Fortunato & Hric, 2016; Porter et al., 2009). One potential future direction that involves community structure and other large-scale network structures is to simultaneously relate individuals' brain data to features of the local structures of their networks, characteristics of their intermediate-scale (i.e., "mesoscale") structures (such as communities), and global network characteristics of a network. As we discussed in the main text, there exist numerous algorithms for identifying communities in a network (Fortunato & Hric, 2016; Porter et al., 2009). There are also methods for characterizing other types of intermediate-scale structures. One example is "core–periphery structure", in which one attempts to detect one or more cores of densely connected nodes, along with sparsely connected peripheral nodes (Csermely et al., 2013; Rombach et al., 2017). Another example is "role structure", in which one attempts to detect similar role structures of nodes (e.g., perhaps the ego networks of graduate students, postdoctoral scholars, and professors have different structural characteristics), regardless of the density of connections (Rossi & Ahmed, 2015). Future research that integrates tools for detection of communities and other mesoscale structures in networks may be fruitful for elucidating how the features of such large-scale structures impact individuals' cognitive processes and behavior.